\documentclass{epl}
\usepackage{amssymb}
\title{
Microwave and millimeter wave spectroscopy in the slightly hole-doped 
ladders of Sr$_{14}$Cu$_{24}$O$_{41}$
}
\author{H. Kitano\inst{1}, R. Inoue\inst{1}, T. Hanaguri\inst{2}, 
A. Maeda\inst{1,3}, N. Motoyama\inst{2},\\ 
M. Takaba\inst{2}, K. Kojima\inst{2}, H. Eisaki\inst{2} \and S. Uchida\inst{2}}
\institute{
  \inst{1} Department of Basic Science, The University of Tokyo - 
  Meguro-ku, Tokyo 153-8902, Japan\\
  \inst{2} Department of Advenced Materials Science, The University of Tokyo - 
  Bunkyo-ku, Tokyo 113-8656, Japan\\
  \inst{3} CREST, Japan Science and Technology Corporation (JST) - 
  Kawaguchi 332-0012, Japan
}

\begin{document}

\maketitle

\begin{abstract}
We have measured the temperature- and frequency dependence of 
the microwave and millimeter wave conductivity $\sigma_1(T,\omega)$ 
along both the ladder ($c$-axis) and the leg ($a$-axis) directions in 
Sr$_{14}$Cu$_{24}$O$_{41}$. 
Below a temperature $T^*$($\sim$170~K), 
we observed a stronger frequency dependence in 
$\sigma_1^c(T,\omega)$ than that in $\sigma_1^a(T,\omega)$, 
forming a small resonance peak developed 
between 30~GHz and 100~GHz. 
We also observed nonlinear dc conduction along 
the $c$-axis at rather low electric fields below $T^*$. 
These results suggest some collective excitation contributes to 
the $c$-axis charge dynamics of the slightly hole-doped ladders of 
Sr$_{14}$Cu$_{24}$O$_{41}$ below $T^*$. 
\end{abstract}

Following a theoretical prediction \cite{Dagotto92} and 
the experimental discovery\cite{Uehara96} of superconductivity 
in the two-leg ladder systems, 
the study of Sr$_{14-x}$Ca$_x$Cu$_{24}$O$_{41}$ 
compounds \cite{Dagotto99} 
with both Cu$_2$O$_3$ ladders and CuO$_2$ chains 
has attracted the attention of many physicists. 
Self-doped holes already exist in this compound, 
since the average valence of Cu is +2.25. 
An early optical measurement revealed that the isovalent substitution 
of Ca for Sr made the self-doped holes redistribute from the CuO$_2$ chains 
to the Cu$_2$O$_3$ ladders, which effectively enables the hole doping 
on the two-leg ladders \cite{Osafune97}. 
They also suggested that the low-energy charge excitation 
($\lesssim$1~eV) was mainly dominated by holes on the Cu$_2$O$_3$ ladders 
rather than on the CuO$_2$ chains, 
because the Cu3$d$-O2$p$ transfer integral of the ladders was larger than 
that of the chains due to the difference of the angle of the Cu-O-Cu bonds. 
Thus, systematic study of the low-energy charge dynamics of 
Sr$_{14-x}$Ca$_x$Cu$_{24}$O$_{41}$ compounds is expected to reveal 
rich physics from the holes on the two-leg ladders, 
which may also provide useful information for understanding 
high-$T_c$ superconductivity in cuprates. 

It is particularly important to study the charge dynamics of 
the parent material Sr$_{14}$Cu$_{24}$O$_{41}$ (estimated 
to have 0.07 holes per ladder Cu site \cite{Osafune97}), 
because it provides a starting point to consider 
strong correlations of holes doped on the two-leg ladders. 
This compound shows semiconductor-like behavior in the dc resistivity 
$\rho_{\rm dc}$ below room temperature \cite{Carter96}. 
Although it was already established that 
a charge-ordered state exists as a dimerization between 
two Cu$^{2+}$ ions in the chain layers below room temperature 
\cite{Matsuda96,Takigawa98,Cox98}, 
some studies on the low-energy charge dynamics of this compound 
\cite{Carter96,Takigawa98} have suggested 
the possibility of a charge ordering in the ladder layers 
below a temperature lower than room temperature. 
In particular, an NMR/NQR measurement \cite{Takigawa98} has suggested that 
the charge dynamics of holes on the ladders which was manifested 
in the quadrupole relaxation at ladder Cu sites below 150~K 
can be explained not by a hopping process of localized holes 
but by a different type of hole motion. 
Thus, in such a possible charge-ordered state of the slightly hole-doped 
ladders, some collective charge excitation related to the strong 
correlated nature of the holes is expected. 

In this paper, 
we describe the charge dynamics along both the ladder 
($c$-axis) and leg ($a$-axis) directions of Sr$_{14}$Cu$_{24}$O$_{41}$ 
in the microwave and millimeter wave regions \cite{preliminary}. 
We observed that a small and narrow conductivity peak developed 
between 30~GHz and 100~GHz below a temperature $T^*$($\sim$170~K) 
in the $c$-axis conductivity $\sigma_1^c(T,\omega)$, 
while we did not observe any sign of it in the $a$-axis conductivity 
$\sigma_1^a(T,\omega)$. 
The resonance-like conductivity peak in the low-energy 
region around 50~GHz ($\sim$0.2~meV) was observed up to moderately high 
temperatures ($\lesssim T^*$$\sim$10~meV). 
Thus, it is difficult to explain this conductivity structure in terms of 
any single particle excitations. 
Rather it is reminiscent of a pinned {\it collective} mode in the density wave 
systems such as charge-density waves (CDW) and spin-density waves (SDW), 
or that in the Wigner crystals. 
We also measured the $c$-axis dc conductivity $\sigma_{\rm dc}^c$ 
as a function of electric field, and found a clear nonlinearity even at 
low electric fields below $T^*$, which is also similar to observations in 
the sliding CDW and SDW states, or that in the Wigner crystals. 
%Furthermore, we observed an anomalous increase in the real part of 
%the complex dielectric constant $\epsilon_1^c(T,\omega)$ along the $c$-axis 
%with decreasing temperature below $T^*$, 
%while $\epsilon_1^a(T,\omega)$ along the $a$-axis continued to 
%decrease with decreasing temperature. 
These results suggest that some collective excitation contributes to 
the peak feature in $\sigma_1^c(\omega)$ and the nonlinear conduction in 
$\sigma_1^c(E)$ below $T^*$. 

\begin{figure}
\onefigure[width=12cm]{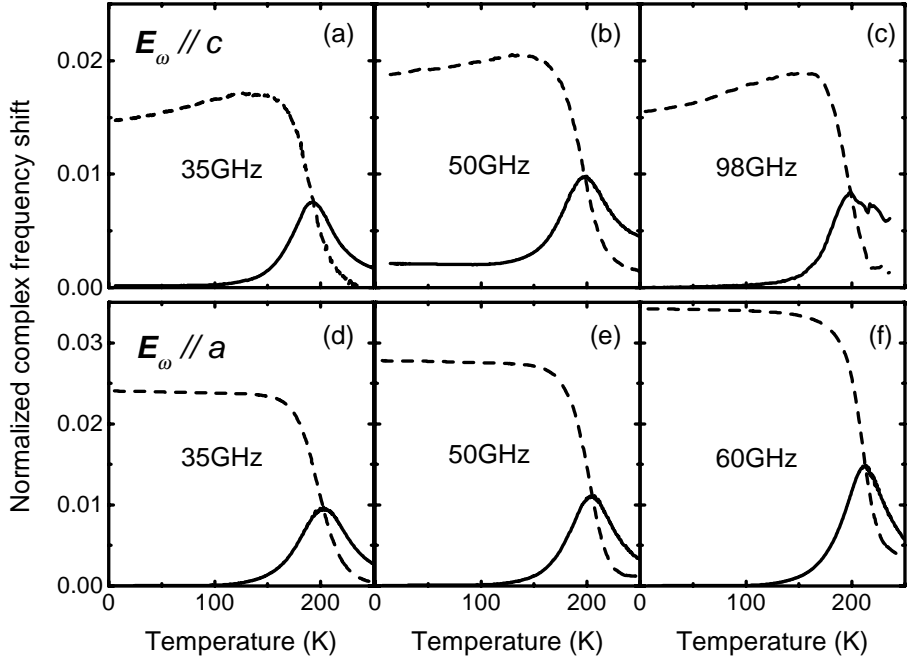}
\caption{
The temperature dependence of 
the real and imaginary parts of the complex frequency shift 
[$\Delta f/f$ (dashed lines) and $\Delta(1/2Q)$ (solid lines), respectively] 
along the $c$-axis (upper panels) and $a$-axis (lower panels). 
They are normalized by the geometrical factor $\gamma/n^2$. 
}
\label{fig1}
\end{figure}

Measurements were made on 9 pieces of 
Sr$_{14}$Cu$_{24}$O$_{41}$ single crystals in 3 batches. 
They were grown by the traveling-solvent-floating-zone (TSFZ) method. 
Typical transport and magnetic properties of these crystals are described 
elsewhere \cite{Motoyama97}. 
All samples were cut into rectangular shapes with various dimensions 
\cite{SampleSize} and microwave and millimeter wave responses were measured by 
the standard cavity perturbation technique \cite{KleinDresselGruner93}. 
To study the frequency dependence of $\sigma_1$, we measured the responses 
at 6 frequencies between 30~GHz and 100~GHz with several Cu (OFC) 
cylindrical cavity resonators operated in the TE$_{011}$ or TE$_{013}$ modes. 
For measurements of the $c$($a$)-axis microwave properties, 
using a sapphire plate fixed on the endplate of the cavity resonator, 
the crystal with the longer dimension along the $c$($a$)-axis was 
placed at a position into the cavity, so that the $\theta$ component of 
the microwave electric field $E_\theta$ was parallel to the $c$($a$)-axis 
of the crystal \cite{Position}. 
To avoid too large a perturbation by the insertion of the crystal 
into the cavity resonator, smaller crystals were used at higher frequencies, 
so that a filling factor $\gamma$, which was proportional to the volume ratio 
of the sample to the cavity, was less than 1$\times$10$^{-4}$. 
In addition, we confirmed that the experimental results were almost 
independent of sample size within experimental error, 
by measuring two or three crystals with different dimensions 
at each frequency (described below). 
%The whole numbers of measurements for 9 crystals at 6 frequencies were 
%above 25. 

As is well known, in the depolarization regime (DPR), 
the real and imaginary parts of 
the complex frequency shift [$\Delta f/f - i\Delta(1/2Q)$], 
(the difference between the presence and absence of a crystal in the cavity), 
are directly related to the complex dielectric constant 
$\epsilon_1+i\epsilon_2$, as \cite{KleinDresselGruner93}

\begin{equation}
\label{eq1}
\begin{array}{l}
\displaystyle\frac{\Delta f}{f}
=\displaystyle\frac{\gamma}{n^2}
              \frac{\left(\epsilon_1-1+\displaystyle\frac{1}{n}\right)}
                   {\left(\epsilon_1-1+\displaystyle\frac{1}{n}\right)^2
                          +\epsilon_2^2}
+C,  \\

\Delta\left(\displaystyle\frac{1}{2Q}\right)
=\displaystyle\frac{\gamma}{n^2}
              \frac{\epsilon_2}
                   {\left(\epsilon_1-1+\displaystyle\frac{1}{n}\right)^2
                          +\epsilon_2^2}, \\
\end{array}
\end{equation}
\noindent
where $n$ is the depolarization factor of the sample 
which can be estimated by an ellipsoidal approximation 
(typically $\sim$0.1 for the measured samples)\cite{Osborn45}, 
and $C$ is an offset of $\Delta f/f$ 
which should be determined experimentally. 

To obtain the conductivity spectra, $\sigma_1(\omega)$, 
from cavity perturbation measurements at several frequencies, 
it is important to estimate carefully the error involved in 
the geometrical factor $\gamma/n^2$ of Eq.~(\ref{eq1}). 
%In the present case, since $\sigma_{dc}$ of this compound shows 
%a strong insulating behavior below 300~K \cite{Motoyama97}, 
%it is impossible to determine $\gamma/n^2$ from measurements 
%at higher temperatures, in a contrast to the case of 
%the quasi-one dimensional (Q1D) CDW materials 
%which often showed the metal-insulator (MI) transition below 300~K. 
Our method was as follows. 
The main source of the error is in the estimation of $n$. 
In the DPR, $C$ is given by $-\gamma/n+\Delta z$, 
where $\Delta z$ is a change related to the mechanical reproducibility 
in opening and closing the cavity for inserting the crystal. 
We found that $|\Delta z|$ could be smaller than 1$\times$10$^{-4}$ 
in our cavity resonators by improving the mechanical reproducibility. 
On the other hand, the constant $C$ could be determined by 
equating $\Delta f/f-C$ and $\Delta(1/2Q)$ at the so-called 
``depolarization peak (DP)" in the DPR, since $\epsilon_2=\epsilon_1-1+1/n$ at 
the DP \cite{KleinDresselGruner93}. 
The determined values of $|C|$ (typically $\sim 1\times$10$^{-3}$) were found 
to be much larger than $|\Delta z|$. 
Thus, it was suggested that $C$ was dominated by $-\gamma/n$ not by 
$\Delta z$. 
This enables us to estimate $n$ from $C$. 
We estimated $n$ by this procedure, and confirmed that it showed a good 
agreement with the value estimated by an ellipsoidal approximation 
within the error range of $\pm20\%$, which corresponded to an error range of 
$\pm40\%$ for estimation of a geometrical factor $\gamma/n^2$ in 
$\Delta (1/2Q)$. 
By measuring two or three crystals with different sizes at each frequency, 
we also found that the magnitude of the DP in $\Delta (1/2Q)$ was almost 
proportional to $\gamma/n^2$ within the error range of $\pm50\%$. 
Thus, we concluded that the errors involved in $\gamma/n^2$ 
were less than $\pm50\%$ in our measurements. 

The conductivity as a function of electric field was 
measured not only by the usual four-probe dc method 
but also by the pulse method using a boxcar averager 
and a digital oscilloscope, in order to check the Joule-heating effect. 
By using rectangular-shaped pulses with a width of 100~$\mu$sec 
and a repetition period of 10~msec, we confirmed that 
the Joule-heating in the pulse measurement was negligibly small below 300~K. 
%Non-heating effect was also checked by the shape of the output pulse 
%displayed on the digital osciloscope (Tektronix, TDS420A). 
%(Stanford Research System, SR250)

Figure 1 shows the typical temperature dependence of 
the real and imaginary parts of the complex frequency shift 
[$\Delta f/f$ (dashed lines) and $\Delta(1/2Q)$ (solid lines), respectively] 
along the $c$-axis (upper panels) and the $a$-axis (lower panels) 
at several frequencies. 
All of the data were normalized by $\gamma/n^2$. 
At all frequencies and along both directions, 
we observed the DPs in $\Delta(1/2Q)$ at around 200~K \cite{Zhai99}. 
By measuring other compounds such as the undoped compound 
La$_{6}$Ca$_{8}$Cu$_{24}$O$_{41}$ and the Ca substituted compounds 
Sr$_{14-x}$Ca$_x$Cu$_{24}$O$_{41}$ ($x$=1,3,12) at 50~GHz, 
we found that the DPs in $\Delta (1/2Q)$ were observed 
only for the compounds with the slightly hole-doped ladders 
($x$=0,1,3), and that the peak position was systematically shifted 
from $\sim$200~K to $\sim$40~K with increasing $x$ from 0 to 3. 
These results suggested that the behavior of $\Delta(1/2Q)$ was strongly 
dependent on the hole concentration on the ladders. 

In addition, as shown in Fig.~1, we found two features related to 
the behavior of $\sigma_1^c$ and $\epsilon_1^c$. 
One feature is that the magnitude of $\Delta(1/2Q)$ for 
$E_\omega\parallel c$ at 50~GHz below $\sim$150~K was much larger than 
at other frequencies, suggesting that the behavior of $\sigma_1^c$ was 
strongly dependent on frequency at around 50~GHz below $\sim$150~K, 
as shown in Fig.~2. 
We emphasize that the frequency-dependent $\sigma_1^c$ is intrinsic 
based on the following reasons. 
(1) The measurements for $E_\omega\parallel c$ on the same 
crystal showed an anomaly only at $\sim$50~GHz, 
and this anomaly was independent of the crystal batch. 
(2) $\Delta(1/2Q)$ for $E_\omega\parallel a$ did not 
show any anomaly at $\sim$50~GHz, suggesting that this anomaly was not 
attributed to the specific property of the 50~GHz cavity resonator. 
(3) This anomaly at $\sim$50~GHz along the $c$-axis was common 
to the compounds with the slightly hole-doped ladders ($x$=0,1,3). 

The other feature is that $\Delta f/f$ for $E_\omega\parallel c$ 
slightly decreased with decreasing temperature below $\sim$150~K 
at all frequencies, while for $E_\omega\parallel a$ it did not show 
any decrease with decreasing temperature. 
We found that this behavior was related to an anomalous increase of 
$\epsilon_1^c$ below $\sim$150~K, as will be shown in Fig.~3. 

Figures 2 and 3 show the temperature dependence of $\sigma_1(T)$ and 
$\epsilon_1(T)$ along both directions at several frequencies. 
They were obtained from $\Delta(1/2Q)$ and $\Delta f/f$ by solving 
Eq.~(\ref{eq1}) inversely for 
$\epsilon_1$ and $\epsilon_2(=4\pi\sigma_1/\omega)$. 
Because Eq.~(\ref{eq1}) is valid only for the DPR, it is important to 
know the region where the concept of the DPR can be applied \cite{Note}. 
%({\it i.e.} $\epsilon_2>\epsilon_1-1+1/n$). 
In fact, if we apply Eq.~(\ref{eq1}) to the regions above 200~K 
(the higher temperature regions above the DP), 
$\sigma_1^c$ above 200~K is found to show apparent metallic 
temperature dependence which is definitely different from 
that of $\sigma_{\rm dc}^c$. 
We found that this disagreement was due to a breakdown of 
the application of Eq.~(\ref{eq1}). 
%indicating that $\sigma_1(T)$ and $\epsilon_1(T)$ above 200~K 
%are no longer analytically obtained from $\Delta(1/2Q)$ and $\Delta f/f$. 
%In such cases, the graphical mapping method is generally needed \cite{Ong77}. 
Therefore, we only discuss the behavior of $\sigma_1(T)$ and $\epsilon_1(T)$ 
up to 200~K below. 

\begin{figure}
\twofigures[width=7cm]{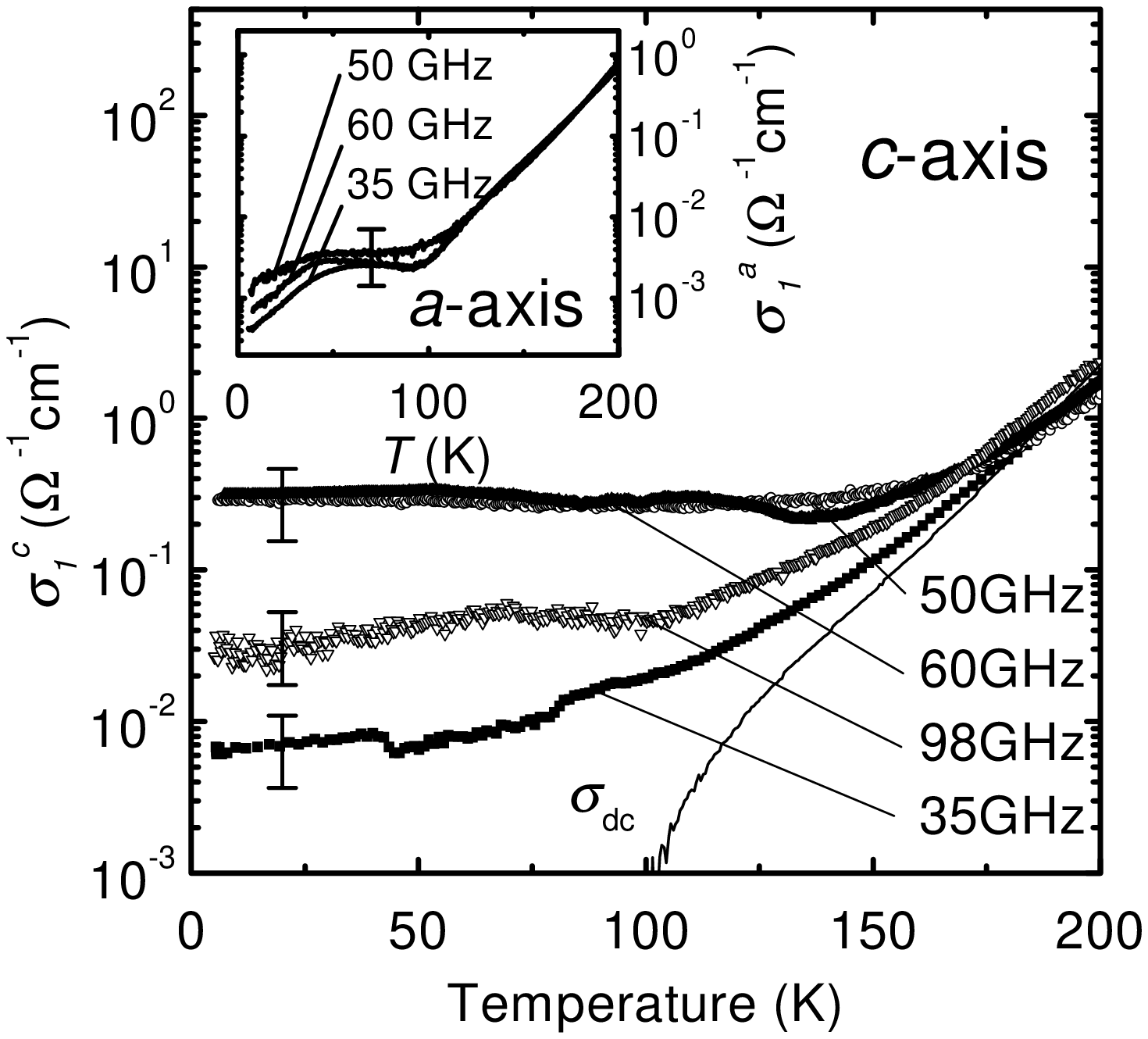}{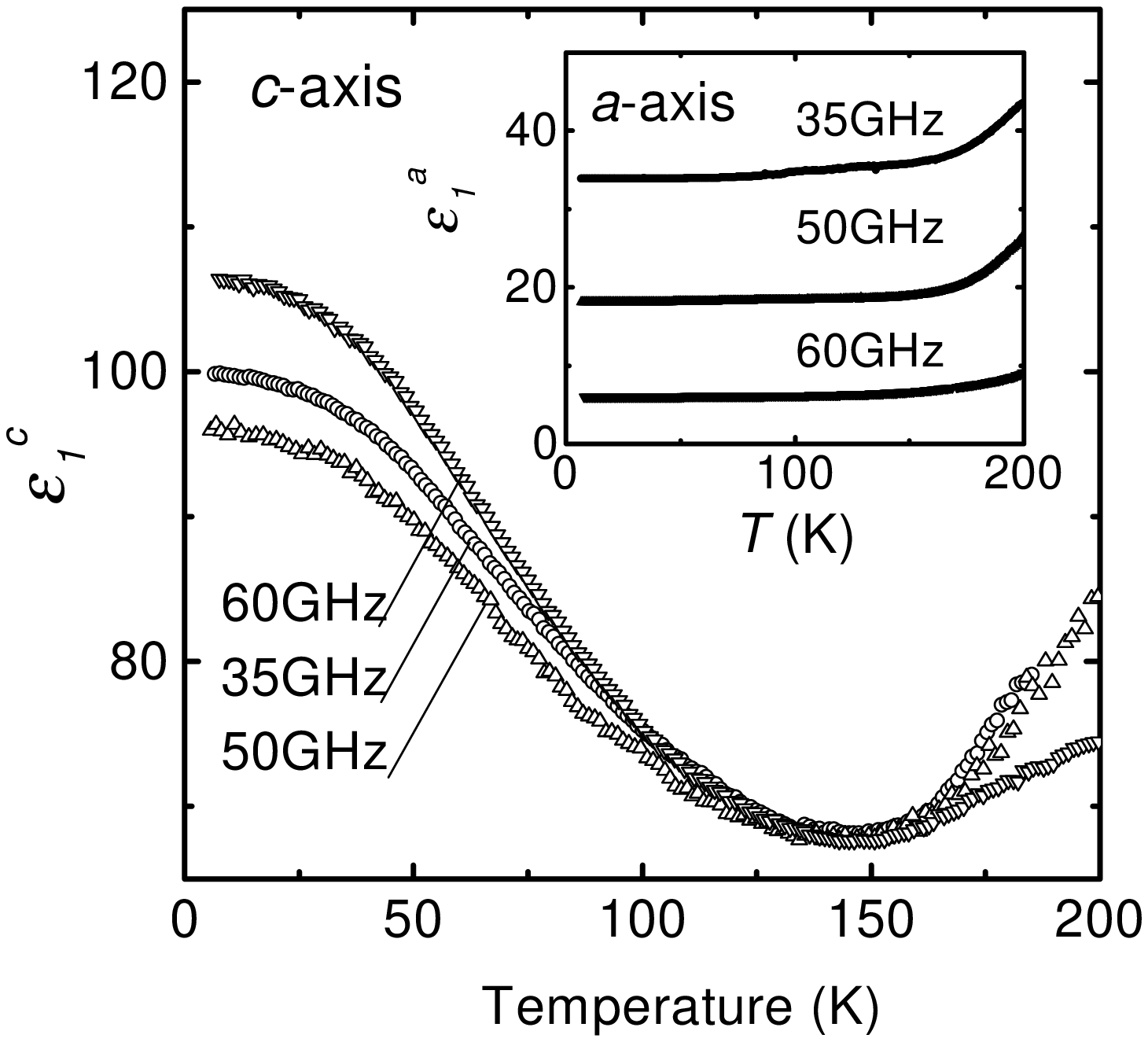}
\caption{
The temperature dependence of $\sigma_1^c(T)$ at several frequencies. 
Inset: 
The temperature dependence of $\sigma_1^a(T)$ at several frequencies. 
}
\label{fig2}
\caption{
The temperature dependence of $\epsilon_1^c(T)$ at several frequencies. 
Inset: 
The temperature dependence of $\epsilon_1^a(T)$ at several frequencies. 
}
\label{fig3}
\end{figure}

As shown in Fig.~2, we found that 
the temperature dependence of $\sigma_1^c(T)$ was strongly frequency-dependent 
below a temperature $T^*(\sim$170~K), while the frequency dependence of 
$\sigma_1^c(T)$ above $T^*$  and $\sigma_1^a(T)$ below 200~K 
was negligibly small within the experimental errors. 
In particular, the magnitude of $\sigma_1^c$ at 50 and 60~GHz was larger than 
those of $\sigma_1^c$ at 35 and 98~GHz by one order of magnitude 
at the lowest temperature. 
This is well above the possible error in the estimation of 
$\gamma/n^2$. 
Thus, as was already discussed, 
it appears that the strongly frequency-dependent $\sigma_1^c(T)$ 
was {\it intrinsic} in the $c$-axis charge dynamics below $T^*$, and 
its frequency dependence implies a resonance-like conductivity peak 
developed below $T^*$. 

The peculiar feature in the $c$-axis charge dynamics below $T^*$ was 
also observed in the temperature dependence of $\epsilon_1^c(T)$, 
as shown in Fig.~3. 
As shown in the inset of Fig.~3, 
$\epsilon_1^a(T)$ continued to decrease with decreasing temperature. 
On the other hand, 
$\epsilon_1^c(T)$ showed a minimum at $\sim$150~K and increased 
with further decreasing temperature at all frequencies. 
Comparing with a recent optical study \cite{Takaba}, 
this increase in $\epsilon_1^c(T)$ suggests some softening of 
a phonon mode which was manifested in the far infrared reflectivity 
as a sharp edge near 38~cm$^{-1}$ below $T^*$. 
On the other hand, the frequency dependence of $\epsilon_1^c(\omega)$ 
appeared to be considerably smaller than that of $\epsilon_1^a(\omega)$ 
both above and below $\sim$150~K within the experimental errors. 
%Qualitatively, it seems that $\epsilon_1^c$ above $\sim$150~K decreases with 
%decreasing frequency like as the behavior of $\epsilon_1^a$, suggesting 
%a dielectric relaxation process is basically dominant along both 
%the $c$- and $a$-axes. 
This is apparently inconsistent with the strongly frequency dependent 
$\sigma_1^c(\omega)$. 
However, since the contribution of the observed peak of $\sigma_1^c$ 
to $\epsilon_1^c$ was expected to be very small (roughly $\sim\pm$10), 
this behavior does not contradict the observed $\sigma_1^c(\omega)$. 
In order to compare $\sigma_1^c(\omega)$ and $\epsilon_1^c(\omega)$ 
quantitatively, improvement in the measurement resolution is needed. 

\begin{figure}
\onefigure[width=10cm]{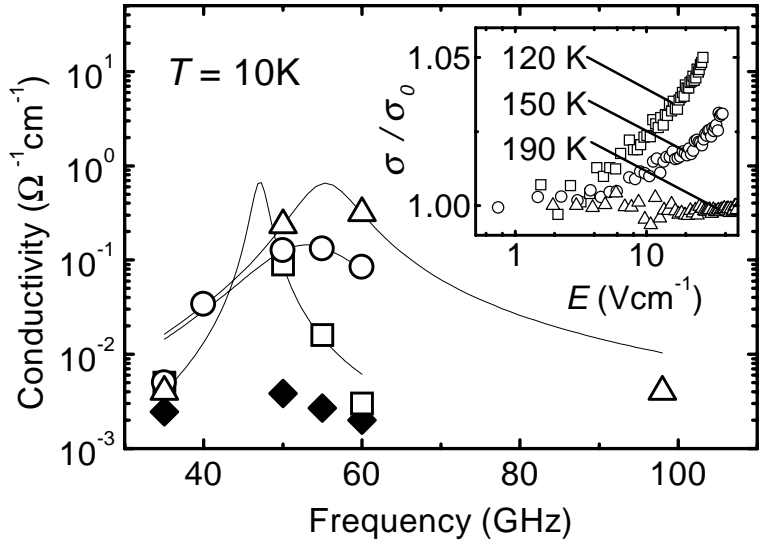}
\caption{
The conductivity spectra of $\sigma_1^c(\omega)$ for 
the crystals from 3 different batches (open marks) 
and $\sigma_1^a(\omega)$ (solid diamonds) at 10~K. 
Soid curves were obtained by fitting to a Lorentzian. 
Inset: 
$\sigma_{\rm dc}^c$ as a function of electric field 
at several temperatures. 
They are normalized by the value in the low-field limit, $\sigma_0$. 
}
\label{fig4}
\end{figure}

The existence of a resonance-like peak in $\sigma_1^c$ was illustrated 
more clearly in the conductivity spectra at 10~K, as shown in Fig.~4. 
It is evident that a relatively small and narrow conductivity peak was 
developed in $\sigma_1^c(\omega)$ between 30~GHz and 100~GHz, 
while the spectrum of $\sigma_1^a(\omega)$ was almost flat. 
We found that the spectra of $\sigma_1^c(\omega)$ could be approximately 
fit to a usual Lorentzian lineshape, although the peak frequency and the width 
of the conductivity peak were slightly dependent on the batch of crystals. 
It is quite important to note that this conductivity peak could be observed 
in the very low energy region ($\sim$0.2~meV) up to moderately high 
temperatures ($\lesssim T^*$$\sim$10~meV). 
If this feature of $\sigma_1^c(\omega)$ in Fig.~4 was attributed to 
some gap feature in the single particle excitation spectrum, it will be 
completely broadened and will be no longer observed at 10~K ($\lesssim$1~meV) 
due to thermal fluctuations. 
In addition, although the inhomogeneity in the crystals may account for 
the presence of any structures in $\sigma_1(\omega)$ in this region, 
it seems to be quite difficult that such an effect accounts for 
the difference betweeen $\sigma_1^c(\omega)$ and $\sigma_1^a(\omega)$. 
Thus, the most possible candidate may be a {\it collective} excitation 
similar to a pinned phason mode in the CDW and SDW states for quasi-one 
dimensional materials \cite{Gruner88,Gruner94}, 
or that in the Wigner crystal observed for the two dimensional hole system of 
GaAs/Al$_{1-x}$Ga$_x$As heterojunction \cite{Li97}. 
Furthermore, the possibility of the collective excitation in $\sigma_1^c$ 
below $T^*$ is also supported by a nonlinear conduction in the electric field 
dependence of $\sigma_{\rm dc}^c$, as shown in the inset of Fig.~4. 
We observed a clear nonlinear conduction at rather low fields below $T^*$. 
This is also similar to observations in sliding CDW and SDW states 
\cite{Gruner88,Gruner94}, or in the Wigner crystal state \cite{Li97}. 
The detailed study of the nonlinear conduction in 
Sr$_{14-x}$Ca$_x$Cu$_{24}$O$_{41}$ compounds will be discussed 
in a different paper \cite{Inoue}. 

Unfortunately, however, we can not specify the origin of the possible 
collective excitation at present. We only speculate that such collective mode 
corresponds to the dynamics of a possible charge-ordered state on the ladders. 
For the density wave systems, it is well known that
the pinning frequency $\omega_0$ of the pinned collective mode is 
related to the electric threshold field for the onset of the nonlinear 
conduction, $E_0$, by 
\begin{equation}
\label{eq2}
eE_0\approx m^*\omega_0^2\lambda, 
\end{equation}
\noindent
where, $m^*$ is the effective mass. 
$\lambda$ is the distance necessary to induce translational motion 
of the collective mode by the electric field. 
For the present collective mode, our results in Fig.~4 suggested that 
$\omega_0/2\pi\sim$50~GHz, and the detailed study of the nonlinear conduction 
\cite{Inoue} suggested that $E_0$ was presumably between 0.1 and 1~V/cm. 
Thus, if $\lambda$ is assumed to be similar to the Cu-Cu distance 
along the $c$-axis on the ladders ($\sim$3.95~\AA), 
we can estimate that $m^*\approx m_e$ ($m_e$ is a free electron mass) 
from Eq.~(\ref{eq2}), suggesting a possible charge-ordered state 
without a lattice distortion or with a negligibly small lattice distortion. 
Since no enhancement in $m^*$ strongly suggests the importance of 
the electron-electron interaction, it is speculated that the strongly 
correlated nature of doped-holes on the ladders is essential. 

On the other hand, we found that the spectral weight (SW) of 
the observed conductivity peak was significantly small. 
In fact, by using $n_h\sim$0.07 as the hole density per ladder-Cu site 
\cite{Osafune97}, the SW determined by a fit to a usual Lorentzian 
(solid curves in Fig.~4) was found to be smaller by 
at least six orders of magnitude than the SW expected for 
the above effective mass ($m^*\approx m_e$). 
A similar reduced SW of the collective mode was also reported in 
the pinned SDW mode in the Bechgaard salts \cite{Gruner94}, 
which remains a serious puzzle. 
Since the electronic correlation is extremely important 
both in the SDW systems and the present ladder systems, 
the reduced SW may be a common important issue for the motion of 
the collective mode in strongly correlated systems. 

%The strongly reduced SW of the collective mode and the small contribution 
%($\sim$5~\%) of the nonlinear conduction to $\sigma_{\rm dc}^c$ rather 
%imply that the amount of holes taking part in such a collective motion 
%may be considerably smaller than those in the conventional CDW state 
%due to the Fermi surface nesting.
%Therefore, we speculate that the observed collective mode has 
%a different origin from the pinned CDW mode. 
%Perhaps, the origin of this collective mode may be related to 
%the strongly correlated nature of the slightly hole-doped ladders. 

In conclusion, by detailed investigation of $\sigma_1(T,\omega)$ and 
$\epsilon_1(T,\omega)$ along both the $c$- and $a$-axes, 
we concluded that some collective excitation probably contributed to 
the $c$-axis charge dynamics in the slightly hole-doped ladders 
of Sr$_{14}$Cu$_{24}$O$_{41}$ below $\sim$170~K. 
Although we cannot specify the origin of the collective mode at present, 
we speculate that it is closely related to a possible charge-ordered state 
of the very small amount of holes doped on ladders, 
and may be characteristic of strongly correlated materials. 

We thank T. Giamarchi for helpful discussions, 
and D. G. Steel for a critical reading of the manuscript. 
This work was partly supported by the Grant-in-Aid for Scientific Research 
from the Ministry of Education, Science, Sports and Culture of Japan.


\begin{thebibliography}{0}
\bibitem{SampleSize}
  Typical dimensions of crystals were 
  0.6$\times$0.3$\times$0.2~mm$^3$ for larger crystals and 
  0.3$\times$0.2$\times$0.1~mm$^3$ for smaller crystals.
\bibitem{Position}
  We have carefully selected the position ($r\sim R/2$, $z=L/2$, 
  where $R$, $L$ are the radius and height of the cavity, respectively) 
  in order to exclude the contribution of the $r$ component of 
  the microwave magnetic field $H_r$.
\bibitem{Note}
  Note that both $\epsilon_1$ and $\epsilon_2$ contribute to 
  the estimation of the skin depths in the present case.
\end{thebibliography}
\end{document}